\documentclass[%
 reprint,
 amsmath,amssymb,
 aps,
]{revtex4-2}

\usepackage{graphicx}
\usepackage{dcolumn}
\usepackage{bm}
\usepackage[
    markup=default, 
    color=true
]{changes}

\begin{document}

\preprint{APS/123-QED}

\thanks{A footnote to the article title}%

\title{Fast CZ Gate via Energy-Level Engineering in Superconducting Qubits with a Tunable Coupler}

\author{Benzheng Yuan$^{1}$}
\email{Benzhengyuan@outlook.com}

\author{Chaojie Zhang$^{1}$}

\author{Chuanbing Han$^{1}$}

\author{Shuya Wang$^{1}$}

\author{Peng Xu$^{1}$}

\author{Huihui Sun$^{1}$}

\author{Qing Mu$^{1}$}

\author{Lixin Wang$^{1}$}

\author{Bo Zhao$^{1}$}
 
\author{Weilong Wang$^{1}$}
\altaffiliation{Corresponding Author}
\email{wangwl19888@163.com}

\author{Zheng Shan$^{1}$}
\altaffiliation{Corresponding Author}
\email{shanzhengzz@163.com}

\address{%
$^{1}$Laboratory for Advanced Computing and Intelligence Engineering, Information Engineering University, Zhengzhou 450001, Henan, China
}

\date{\today}

\date{\today}

\begin{abstract}
In superconducting quantum circuits, decoherence errors in qubits constitute a critical factor limiting quantum gate performance. To mitigate decoherence-induced gate infidelity, rapid implementation of quantum gates is essential. Here we propose a scheme for rapid controlled-Z (CZ) gate implementation through energy-level engineering, which leverages Rabi oscillations between the $\left|11\right\rangle$ state and the non-computational state in a tunable-coupler architecture. Numerical simulations achieved a $\mathrm{22~ns}$ nonadiabatic CZ gate with fidelity over $99.99\%$. We further investigated the performance of the CZ gate in the presence of anharmonicity offsets. The results demonstrate that a high-fidelity CZ gate with an error rate below $10^{-4}$ remains achievable even with finite anharmonicity variations. Furthermore, the detrimental impact of spectator qubits in different quantum states on the fidelity of CZ gate is effectively suppressed by incorporating a tunable coupler. This scheme exhibits potential for extending the circuit execution depth constrained by coherence time limitations.

\end{abstract}

\maketitle


\section{\label{sec:level1}Introduction}

Achieving high-fidelity quantum gates is a critical challenge in the realization of fault-tolerant quantum computation \cite{acharya2024quantum,klimov2024optimizing}. Decoherence errors constitute the dominant error source in the superconducting quantum processor. To mitigate decoherence errors in superconducting qubits, extensive research has focused on advancing qubit hardware design and material processing techniques \cite{gyenis2021experimental, gyenis2021moving, siddiqi2021engineering, hassani2023inductively, somoroff2023millisecond, place2021new, wang2022towards, ganjam2024surpassing,verjauw2022path, kim2021enhanced, bland20252d}. On the other hand, rapid gate implementations are essential to suppress decoherence-induced infidelity, and also enable deeper quantum circuits to be executed within the finite coherence times of qubits. These advancements are particularly critical for widely deployed quantum gates such as the CZ gate.
Current physical implementations of CZ gates in superconducting circuits can be categorized into two paradigms based on their evolution mechanisms: adiabatic \cite{martinis2014fast, xu2020high, chu2021coupler} and nonadiabatic approaches \cite{sung2021realization, barends2019diabatic, li2019realisation}. The adiabatic scheme relies on the cross-Kerr type ZZ interaction between two qubits, achieving phase accumulation on $|11\rangle$ state through parametric modulation. This approach typically employs either frequency tuning of qubits or couplers to achieve phase accumulation \cite{chu2021coupler, goto2022double, li2024realization}, or microwave pulse-driven mediation of ZZ interactions \cite{huang2024fast, jiang2025microwave}. Nonadiabatic CZ gates utilize Rabi oscillations between $|11\rangle$ and either $|02\rangle$ or $|20\rangle$ states \cite{sung2021realization, barends2019diabatic, li2019realisation}. In summary, the implementation of CZ gates typically utilizes the interaction between two states within the double excitation mainfold.

To overcome the speed limitations of the conventional CZ gate, we introduce an energy-level engineering strategy designed to enhance the effective qubit-qubit interaction. An enhanced effective coupling strength of $2g$ is achieved by engineering a resonance between the $|11\rangle$ state and the $(|20\rangle+|02\rangle)/\sqrt{2}$ superposition state, compared to the conventional $\sqrt{2g}$. The resonance condition is satisfied when the energy levels $E_{11}=E_{20}=E_{02}$, enabling high-speed gate operation. This spectral alignment requires a pair of qubits with opposing anharmonicities and a frequency separation equal to the magnitude of the anharmonicity. Although similar spectral conditions have been explored in ZZ-suppression architectures, for example, in hybrid Transmon–CSFQ (capacitively shunted flux qubit) systems \cite{zhao2020high, ku2020suppression, xu2021zz}, their practical implementation via direct capacitive coupling is hindered by a fundamental scalability issue: the frequency tuning necessary for gate operation often induces unintended crosstalk with spectator qubits in densely spaced frequency spectra. These unwanted interactions significantly compromise gate fidelity in multi-qubit settings \cite{krinner2020benchmarking,zhao2022quantum}. Consequently, conventional capacitive coupling methods face inherent limitations when extended to one- or two-dimensional qubit arrays.

To address this fundamental constraint while maintaining high-speed gate operation, we introduce a tunable coupler architecture that enables frequency-independent coupling control. The proposed design incorporates a non-resonant pair consisting of a Transmon and an inductively shunted transmon (IST), the latter providing positive anharmonicity at the operating point \cite{fasciati2024complementing}, coupled via a flux-tunable transmon coupler \cite{yan2018tunable}. By leveraging Rabi oscillations between the $|110\rangle$ state and the non-computational state, we achieve an enhanced effective Rabi frequency of $2g$, thereby reducing the duration of CZ gate to $\pi/(2g)$ with an infidelity below $10^{-4}$. To evaluate robustness against fabrication variations, we further investigate the performance of CZ gate under anharmonicity offsets ($\delta\neq0$). The numerical results confirm that the CZ gate maintains an error rate below $10^{-4}$ even with finite anharmonicity deviations. Moreover, by incorporating the tunable coupler, the architecture effectively suppresses crosstalk from spectator qubits in arbitrary quantum states, ensuring high fidelity in multi-qubit environments.

 The paper is structured as follows: Section II presents a theoretical analysis of the Hamiltonian governing the circuit system. Section III explains the operational principles of rapid CZ gates, elucidating the physical origins of their accelerated operation speeds. Section IV delivers comprehensive numerical results. Finally, the work is summarized in the Section V.

\section{Physical Model}
\begin{figure}[h]
\includegraphics[width=0.9\linewidth]{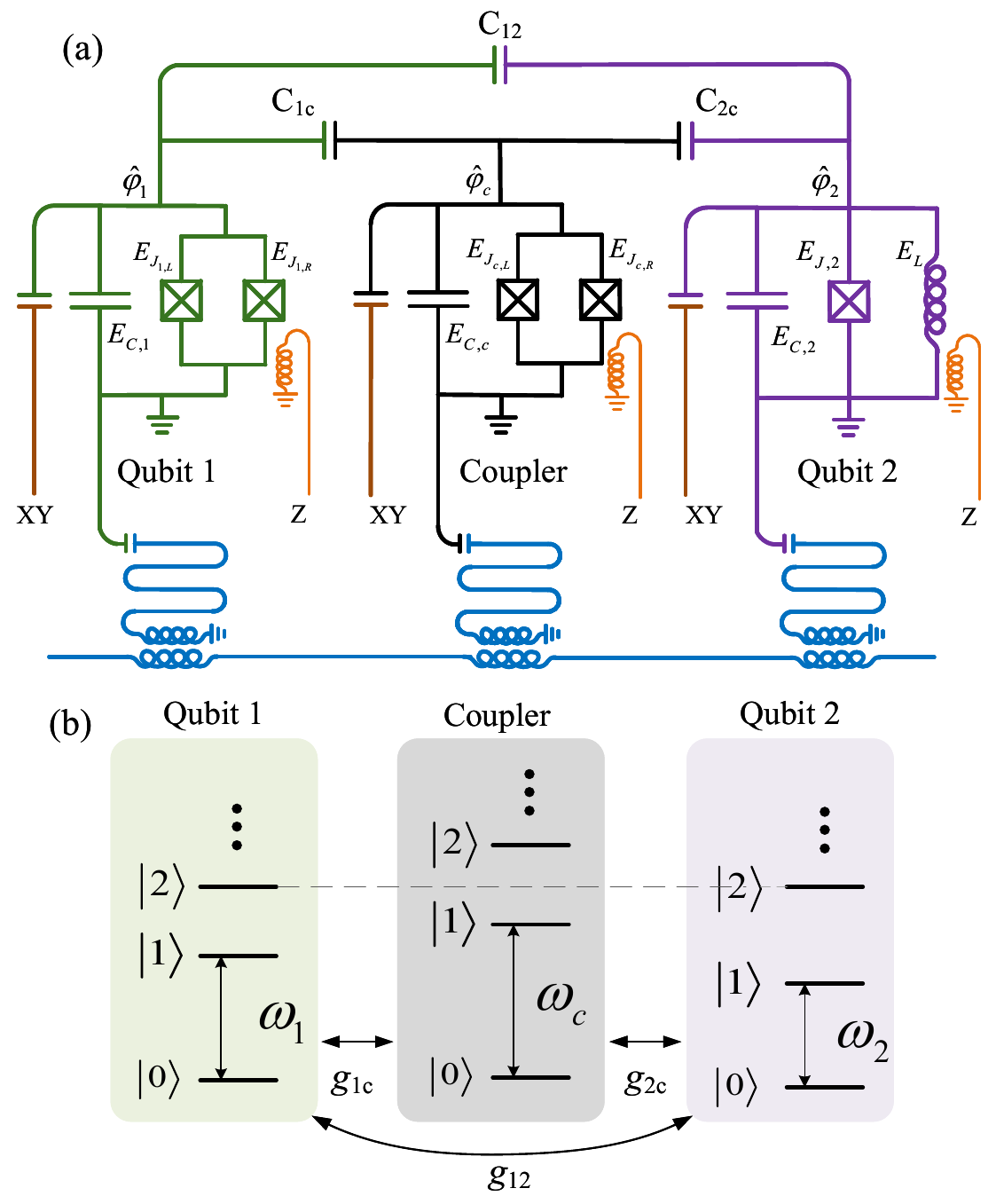}
\caption{\label{Circuit} (a) Schematic of the superconducting circuit featuring a Transmon and an IST qubit coupled via a tunable coupler. (b) Corresponding energy-level diagram demonstrating the condition $E_{110}=E_{200}=E_{020}$ (subscripts denote Q1, Q2, coupler) required for a fast CZ gate. The coupling strengths are denoted by $g_{12}/2\pi=10~\mathrm{MHz}$ (qubit-qubit) and $g_{ic}/2\pi=100~\mathrm{MHz}$ (qubit-coupler, $i=1,2$).}
\end{figure}

Tunable coupling has been demonstrated as a leading physical architecture. We now consider the physical model depicted in Fig.~\ref{Circuit}(a), which consists of a Transmon and an IST coupled via a tunable coupler. The Hamiltonian of system can be described as:
\begin{equation}\begin{aligned}\hat{H}_{s}&=\sum_{i=1,c}4E_{Ci}\hat{n}_{i}^{2}-2E_{Ji}\left|\cos\left(2\pi\phi_{ext}^{i}\right)\right|\cos\hat{\varphi}_{i}\\&+4E_{C2}\hat{n}_{2}^{2}-E_{J2}\cos\left(\hat{\varphi}_{2}+2\pi\phi_{ext}\right)+\frac{1}{2}E_{L}\hat{\varphi}_{2}^{2}\\&+J_{1c}\hat{n}_{1}\hat{n}_{c}+J_{2c}\hat{n}_{2}\hat{n}_{c}+J_{12}\hat{n}_{1}\hat{n}_{2},\end{aligned}\end{equation}
where the indices $i\mathrm{=1,2}$ denote the Transmon and IST modes, respectively, and $c$ lables the transmon coupler mode. $E_C,E_J$ and $E_{L}$ represent charging energy, Josephson energy and inductive energy, respectively. $\phi_{ext}$ is the external flux bias.
The IST is working at half-flux quanta $\begin{pmatrix}\phi_{ext}=0.5\end{pmatrix}$, realizing a single-well plasmonic spectrum with positive anharmonicity ${\alpha\geq0}$ (see Appendix A for details). 
Our protocol relies on an opposite-anharmonicity element whose positive anharmonicity can be engineered to be comparable in magnitude to the negative anharmonicity of the transmon (typically $\sim$200--300~MHz in the parameter range considered here), so as to satisfy the near-degeneracy condition in Fig.~\ref{Circuit}(b). CSFQs can exhibit positive anharmonicity \cite{yan2016flux}; however, reported devices often have anharmonicities of 500--900~MHz, which may exceed the value needed for matching in our architecture and thereby tighten the parameter constraints. In contrast, the IST provides an additional design knob through the inductive shunt, allowing the positive anharmonicity to be engineered below 300~MHz. From an implementation perspective, CSFQs typically employ multiple Josephson junctions with targeted energy ratios, whereas the IST replaces this element with a geometric inductor, reducing circuit complexity and the number of junction parameters that must be controlled. These considerations motivate our choice of the IST as the positive-anharmonicity component in the present work.

We approximate the qubits and the coupler as harmonic modes, i.e., introducing $\hat{\varphi}_i=\hat{\varphi}_{i,z\mathrm{pf}}\left(\hat{a}_i^\dagger+\hat{a}_i\right)$ and $\hat{n}_{i}=i\hat{n}_{i,z\mathrm{pf}}\left(\hat{a}_{i}^{\dagger}-\hat{a}_{i}\right)$, where phase (number) zero-point fluctuation $\hat{n}_{i,z\mathrm{pf}}=1/\left(2\hat{\varphi}_{i,z\mathrm{pf}}\right)$. Thus, the system Hamiltonian can be written as follows:
\begin{equation}\begin{aligned}\hat{H}/\hbar&=\sum_{i=1,2,c}\left(\omega_i\hat{a}_i^\dagger\hat{a}_i+\frac{\alpha_i}{2}\hat{a}_i^\dagger\hat{a}_i^\dagger\hat{a}_i\hat{a}_i\right)\\&+\sum_{i<j}g_{ij}\left(\hat{a}_i^\dagger\hat{a}_j+\hat{a}_i\hat{a}_j^\dagger-\hat{a}_i\hat{a}_j-\hat{a}_i^\dagger\hat{a}_j^\dagger\right),\end{aligned}\end{equation}
where $\hat{a}_{i}^{\dagger}$ and ${{{\hat a}_i}}$ are, respectively, the raising and lowering operators truncated to the lowest four levels (labeled as $\{|0\rangle,|1\rangle,|2\rangle,|3\rangle\}$), defined in the eigenbasis of the corresponding oscillators. $g_{ij}(i,j = 1,2,c)$ denotes the coupling strength. The derivation of the circuit quantization is detailed in Appendix A.

The energy-level structure of each qubit is shown in Fig.~\ref{Circuit}(b). The Hamiltonian under the rotating-wave approximation (RWA) including levels with two excitations for the system in Fig.~\ref{Circuit}(a) can be written as
\begin{equation}\hat{H}=\begin{bmatrix}\omega_{000}&0&0&0&0&0\\0&\omega_{010}&0&g&0&0\\0&0&\omega_{020}&0&\sqrt{2}g_{12}&0\\0&g&0&\omega_{100}&0&0\\0&0&\sqrt{2}g_{12}&0&\omega_{110}&\sqrt{2}g_{12}\\0&0&0&0&\sqrt{2}g_{12}&\omega_{200}\end{bmatrix}\end{equation}
in the $\{|000\rangle,|010\rangle,|020\rangle,|100\rangle,|110\rangle,|200\rangle\}$ basis, where the bare state of the system is denoted by $\left|n_in_jn_c\right\rangle=\left|n_i\right\rangle\otimes\left|n_j\right\rangle\otimes\left|n_c\right\rangle\left(n_i,n_j\in\left\{0,1,2\right\},n_c=0\right)$.

\section{THEORETICAL Analysis FOR CZ GATE IMPLEMENTATION}
\subsection{CZ Gate In the Dispersive Regime}
\begin{figure}[h]
\includegraphics[width=0.9\linewidth]{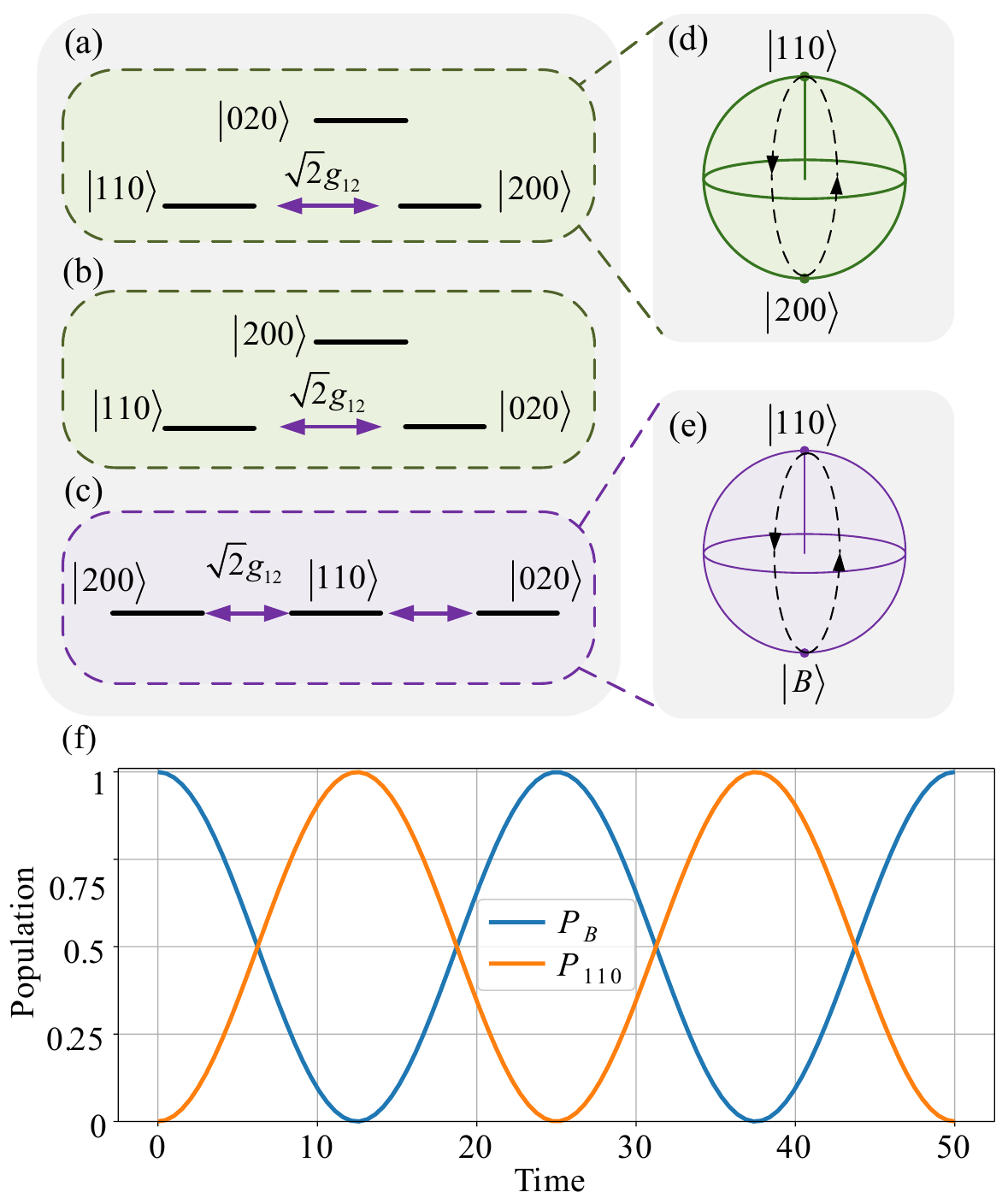}
\caption{\label{Level_2} Mechanisms for conventional and accelerated CZ gate implementation. (a, b) Energy-level diagrams and coupling configurations for conventional CZ gates mediated by the  $|110\rangle-|020\rangle$ and $|110\rangle-|200\rangle$ interaction, respectively. (c) Engineered energy-level structure satisfying  $\omega_{i}+\alpha_{i}=\omega_{j}~(i,j\in\{1,2\})$. (d) Bloch sphere representation of the state evolution under conventional couplings. (e) Bloch sphere representation of the state evolution between $\left|110\right\rangle$ and $\left|B\right\rangle=\left(\left|200\right\rangle+\left|020\right\rangle\right)/\sqrt{2}$. (f) Coherent oscillation between $\left|110\right\rangle$  and $\left|B\right\rangle$under the effective Hamiltonian.}
\end{figure}

The CZ gate is a fundamental two-qubit phase gate widely used in superconducting quantum circuits. Conventional CZ gate implementations utilize the $|110\rangle-|020\rangle$ or $|110\rangle-|200\rangle$ interactions, with the associated energy-level structures and coupling mechanisms depicted in Fig.~\ref{Level_2}(a) and Fig.~\ref{Level_2}(b). Under this configuration, the system Hamiltonian in the rotating frame can be expressed as:
\begin{equation}H^{*}=\sqrt{2}g_{12}\left(\left|110\right\rangle\langle200|+\left|200\right\rangle\langle110|\right)\end{equation}
The resulting evolution, described by the operator $U = \exp(-iH^* t)$, can be visualized on the Bloch sphere as shown in Fig.~\ref{Level_2}(d). The CZ gate is realized when the evolution time reaches $t = \pi /(\sqrt{2}g_{12})$. It is evident that the gate duration in this scheme is fundamentally limited by the underlying coupling mechanism. To overcome this speed constraint, we now introduce an alternative approach for rapidly implementing the CZ gate through deliberate engineering of the energy-level structure.

\begin{table}[b]
\caption{\label{tab:table1}%
Qubit Parameters for Implementing the Fast CZ Gate. These parameters can be derived from the Hamiltonian in Appendix A.
}
\begin{ruledtabular}
\begin{tabular}{lcc}
\textrm{Qubit}&
\textrm{Transmon}&
\textrm{IST}\\
\colrule
Frequency(GHz)     & 4.50  & 4.25\\
Anharmonicity(MHz) & -250  & 250\\
$E_{L}$(GHz)       & -& 24.5\\
$E_{J}$(GHz)   & 14.0& 14.9\\
$E_{C}$(MHz)   & 221& 228\\
\end{tabular}
\end{ruledtabular}
\end{table}

To rapidly implement the CZ gate, the parameters of the qubits can be set as $\omega_{i}+\alpha_{i}=\omega_{j}\left(i,j\in\{1,2\}\right)$, the energy level structure and coupling are shown in Fig.~\ref{Level_2}(c). The opposite anharmonicity required between the two qubits in this scheme is satisfied through the Transmon-IST coupling configuration described in Section II, with the complete parameter set provided in Table \ref{tab:table1}. Under these parameter conditions, after the rotating frame with transform operator $V = \exp \left[ { - i({\omega _{200}}\left| {200} \right\rangle \left\langle {200} \right| + {\omega _{020}}\left| {020} \right\rangle \left\langle {020} \right| + {\omega _{110}}\left| {110} \right\rangle \left\langle {110} \right|)t} \right]$. Then Eq. (3) can be written as 
\begin{equation}\begin{aligned}\hat{H^{\prime}}&=\sqrt{2}g_{12}\left(\left|110\right\rangle\left\langle020\right|+\left|020\right\rangle\left\langle110\right|\right)\\&+\sqrt{2}g_{12}\left(\left|110\right\rangle\left\langle200\right|+\left|200\right\rangle\left\langle110\right|\right)\end{aligned}\end{equation}
with the introduction of state $\left|B\right\rangle$. The system Hamiltonian becomes \begin{equation}\hat{H}^{\prime}=2g_{12}\left(\left|110\right\rangle\left\langle B\right|+\mathrm{H.c.}\right),\end{equation}
where the state $\left|B\right\rangle=\left(\left|200\right\rangle+\left|020\right\rangle\right)/\sqrt{2}$. The evolution between the $\left|110\right\rangle$ and $\left|B\right\rangle$ states, governed by an interaction strength of $2g_{12}$, is illustrated on the Bloch sphere in Fig.~\ref{Level_2}(e). The enhancement of the coupling strength from $\sqrt{2}g_{12}$ to $2g_{12}$ represents not merely a numerical improvement, but a fundamental optimization of the underlying physical mechanism that directly enables the accelerated performance of the CZ gate.

In the basis $\{|110\rangle,|B\rangle\}$, we can map $\begin{pmatrix}|110\rangle\langle B|+\mathrm{H.c.}\end{pmatrix}$ to Pauli matrix $\sigma_{x}$. Based on this, when the evolution time $t=\pi/2g_{12}$, the $\left|11\right\rangle$ state evolves to $-\left|11\right\rangle$. The dynamics generated by this Hamiltonian is shown in Fig.~\ref{Level_2}(f). A coherent oscillation between the $\left|11\right\rangle$ state and the $\left|B\right\rangle$ state can be realized. A full oscillation cycle completes the phase accumulation required for the CZ gate. In the following, we numerically simulate the performance of this scheme under realistic experimental conditions.

\subsection{CZ Gate In the Non-dispersive Regime}

In the weakly dispersive regime or even in the non-dispersive regime, when the detuning between the qubit frequencies and the coupler frequency is varied during the CZ gate operation, not only the states within the qubit double-excitation subspace  $\{|200\rangle,|110\rangle,|020\rangle\}$ participate in the dynamics, but also the coupler-excited states $\{|101\rangle,|011\rangle,|002\rangle\}$ are involved. Nevertheless, the system can still evolve from $\{|110\rangle\}$ to an eigenstate of the coupled Hamiltonian. During this evolution, the eigenstate acquires a well-defined dynamical phase. By appropriately tuning the system back to its initial configuration, the state can return to $|110\rangle$,  accumulating a total phase of $\pi$.

In the double-excitation subspace, with the basis ordered as $\{|200\rangle,|110\rangle,|020\rangle,|101\rangle,|011\rangle,|002\rangle\}$, we diagonalize the Hamiltonian 
$H$ to obtain the eigenvalues $\lambda_{i}$ and the corresponding eigenstates $|E_i\rangle~(i=1,\ldots,6)$. The Hamiltonian can be expressed in terms of these eigenstates as:
\begin{equation}H=\sum_{i=1}^6\lambda_i|E_i\rangle\langle E_i|.\end{equation}
The time evolution operator $U(t)$ can be written as:
\begin{equation}U(t)=\sum_{i=1}^6e^{-i\lambda_it}|E_i\rangle\langle E_i|.\end{equation}
After evolution time $t$, the $\left|110\right\rangle$ state becomes:
\begin{equation}|\psi(t)\rangle=U(t)|110\rangle=\sum_{i=1}^6c_ie^{-i\lambda_it}|E_i\rangle,\end{equation}
where $c_{i}=\langle E_{i}|110\rangle$. We focus on the projection of the evolved state onto $\left|110\right\rangle$, i.e., the probability amplitude $\langle110|\psi(t)\rangle$. This is described as follows:
\begin{equation}\langle110|\psi(t)\rangle=\sum_{i=1}^6|c_i|^2e^{-i\lambda_it}.\end{equation}
To realize a CZ gate, the evolved state must satisfy $|\psi(t)\rangle=-|110\rangle$. This condition is equivalent to demanding that the overlap amplitude equals to $-1$: $\sum_{i=1}^6|c_i|^2e^{-i\lambda_it}=-1.$

\begin{figure}[h]
\includegraphics[width=1\linewidth]{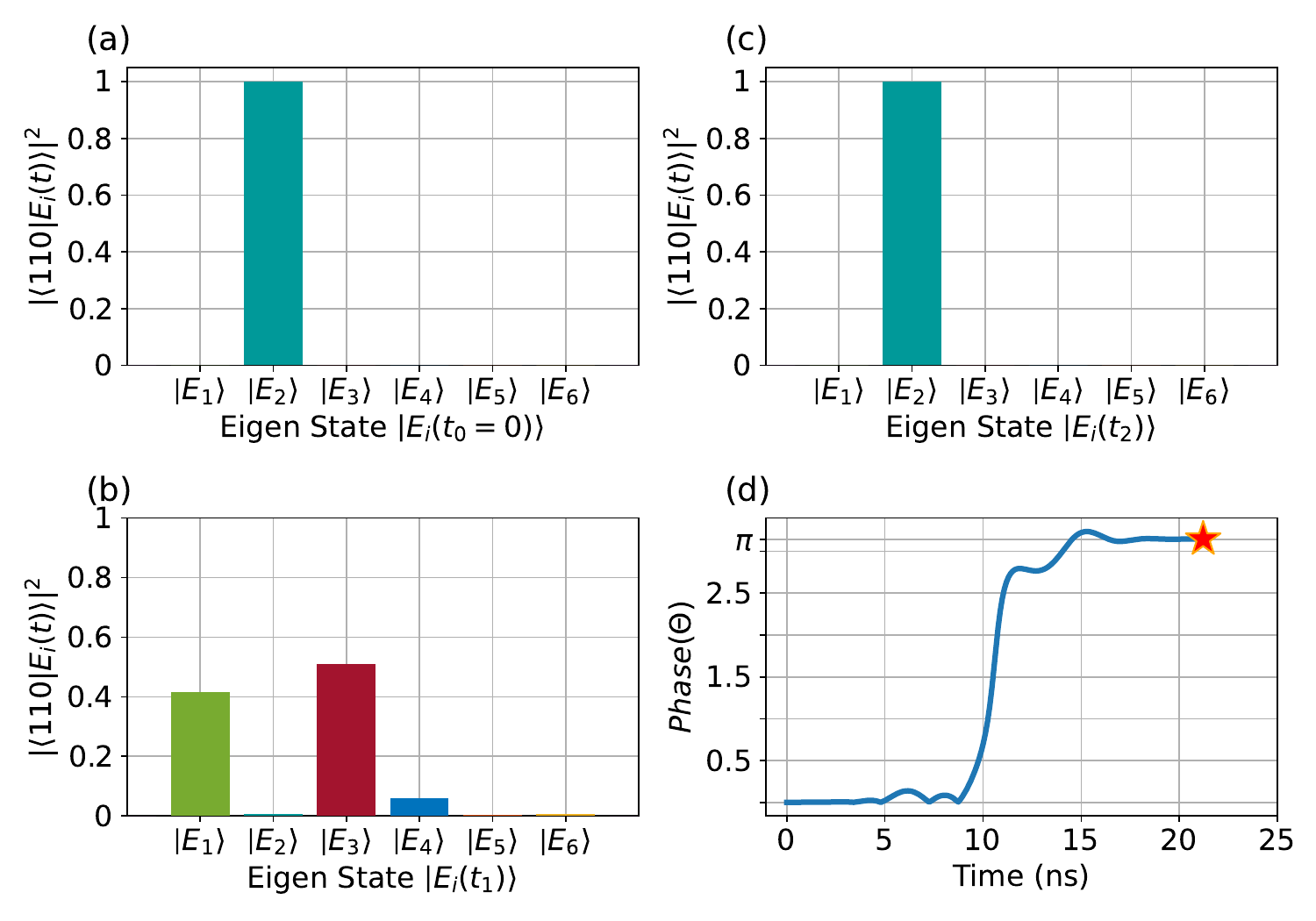}
\caption{\label{overlap_and_phase_t} 
Overlap of the computational basis state $\left|110\right\rangle$ with the Hamiltonian eigenstates and the associated dynamical phase accumulation during time evolution. (a) At the initial time $t_0=0$, the prepared state $\left|110\right\rangle$ is decomposed in the eigenbasis, showing its initial overlap with each eigenstate. (b) At an intermediate time $t_1$, unitary evolution redistributes the state population across multiple eigenstates, yielding a superposition with time-dependent relative phases. (c) By the final time $t_2$, the system revives to $\left|110\right\rangle$ and acquires a net phase of $\pi$ relative to the initial state. (d) Continuous evolution of the absolute value of the accumulated phase  over the full interval [$t_0$,$t_2$], illustrating its smooth buildup to $\pi$ at $t_2$.}

\end{figure}

Based on Eq.~(10), we analyze the time-dependent overlap between the state $|110\rangle$ and the Hamiltonian eigenstates. Specifically, referring to the control waveform shown in Fig.~\ref{overlap_and_phase_t}(a), we select three representative time instants to illustrate the state decomposition. At the initial time $t_0 = 0$, the $|110\rangle$ state exhibits near-unity overlap with eigenstate \(|E_2\rangle\), as shown in Fig.~\ref{overlap_and_phase_t}(a). At the intermediate time \(t_1\), unitary evolution drives the system to a coherent superposition with appreciable weight on several eigenstates as shown in Fig.~\ref{overlap_and_phase_t}(b). By the final time \(t_2\), the state has accumulated a total phase of \(\pi\), and thus revives to $|110\rangle$. The continuous accumulation of this phase over the interval [$t_0$, $t_2$] is shown in Fig.~\ref{overlap_and_phase_t}(d).

\section{Performance of CZ Gate}
To evaluate the performance of rapid CZ gates in this architecture, we perform numerical simulations based on the full Hamiltonian in Eq.(2). The average gate fidelity is then calculated using the standardized metric~\cite{pedersen2007fidelity},

\begin{equation}\overline{F}=\frac{\left|\mathrm{Tr}\left(\hat{U}_{\mathrm{id}}^\dagger\hat{U}\right)\right|^2+\mathrm{Tr}\left(\hat{U}^\dagger\hat{U}\right)}{d(d+1)},\end{equation}
where $d=2^n$ for an n-qubit system, $\hat{U}_\mathrm{id}$ represents the ideal unitary operation, and $\hat{U}$ denotes the numerically simulated operation. All numerical simulations are performed using the QuTiP (Quantum Toolbox in Python) package~\cite{lambert2024qutip}.

\begin{figure}[h]
\includegraphics[width=1\linewidth]{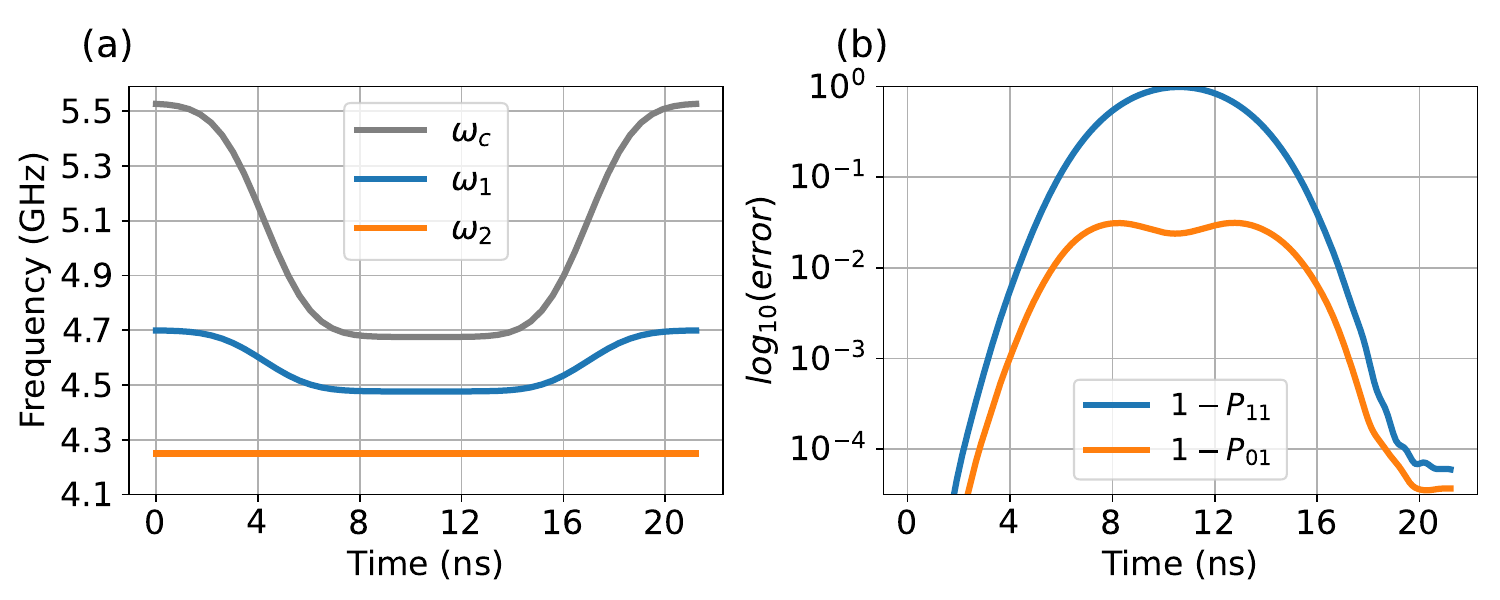}
\caption{\label{wave_pop_0} (a) Frequency pulse scheme for realizing a fast CZ gate by simultaneously tuning the frequencies of the Transmon and the coupler to shift the system from the idle point to the work point. The anharmonicities are set as $-\alpha_1=\alpha_2=2\pi\times250~\mathrm{MHz}$. (b) Simulated leakage and swap errors of the fast CZ gate as functions of the control parameters defined in (a).}
\end{figure}

The main idea of implementing the CZ gate is as follows. In practical circuit design, the qubit frequencies and anharmonicity can be engineered such that $\omega_{i}+\alpha_{i}=\omega_{j}\quad\left (i,j\in\{1,2\}\right)$. Under ideal conditions (assuming ideal fabrication conditions), the qubit frequencies can be set using fixed-frequency qubits to satisfy the energy-level requirement. However, considering the frequency drift and fabrication variability inherent, one Transmon is designed to be frequency tunable. By simultaneously adjusting the frequencies of both the coupler and the Transmon from their respective idle points ($\omega_c^\mathrm{off}$ and $\omega_a^\mathrm{off}$) to their working points, according to the time-dependent pulse function as shown in Fig.~\ref{wave_pop_0}(a). The interqubit coupling $\left|110\right\rangle\leftrightarrow\left|B\right\rangle$ can be activated, enabling a fast CZ gate. 

The frequency pulse profile that we use throughout this work is
\begin{equation}\begin{aligned}\omega_{i}(t)&=\omega_\mathrm{off}+\frac{\omega_\mathrm{on}-\omega_\mathrm{off}}{2}[\operatorname{Erf}(\frac{t-\frac{1}{2}t_{\mathrm{ramp}}}{\sqrt{2}\sigma})\\&-\operatorname{Erf}(\frac{t-t_{\mathrm{gate}}+\frac{1}{2}t_{\mathrm{ramp}}}{\sqrt{2}\sigma})],\end{aligned}\end{equation}
where $\omega_\mathrm{off}$ and $\omega_\mathrm{on}$ denote the frequency of  the qubit/coupler at the parking point and the interaction point. $t_{\mathrm{ramp}}=4\sqrt{2}\sigma$ is the ramp time, $t_{\mathrm{gate}}$ denotes the total gate time for the CZ gate. $t_{\mathrm{hold}}=t_{\mathrm{gate}}-t_{\mathrm{ramp}}$ denotes the hold time that is defined as the time interval between the midpoints of the ramps.

By simultaneously optimizing the operating frequencies of the coupler $\omega _{{\rm{on}}}^c$ and the qubit $\omega_\mathrm{on}^q$ with the pulse sequence illustrated in Fig.~\ref{wave_pop_0}(a) and initializing the system in states $\left|11\right\rangle^{\prime}$ and $\left|01\right\rangle^{\prime}$, we calculate the leakage error, defined as ${\varepsilon _{leak}} = 1 - {P_{11}}$,  and the swap error, defined as ${\varepsilon _{{\rm{swap}}}} = 1 - {P_{01}}({P_{10}})$, as a function of the hold time. Here, ${P_{ij}}$ denotes the population in the state $\left| {ij} \right\rangle $, although the label omits the coupler; the full physical system is a three-mode system that includes it. The gate operation is optimized to minimize leakage, which demonstrates that a CZ gate with fidelity exceeding 99.99\% can be achieved within a hold time of  22 ns, while the leakage and swap errors are suppressed below $10^{-4}$, as show in Fig.~\ref{wave_pop_0}(b).

\begin{figure}[h]
\includegraphics[width=1\linewidth]{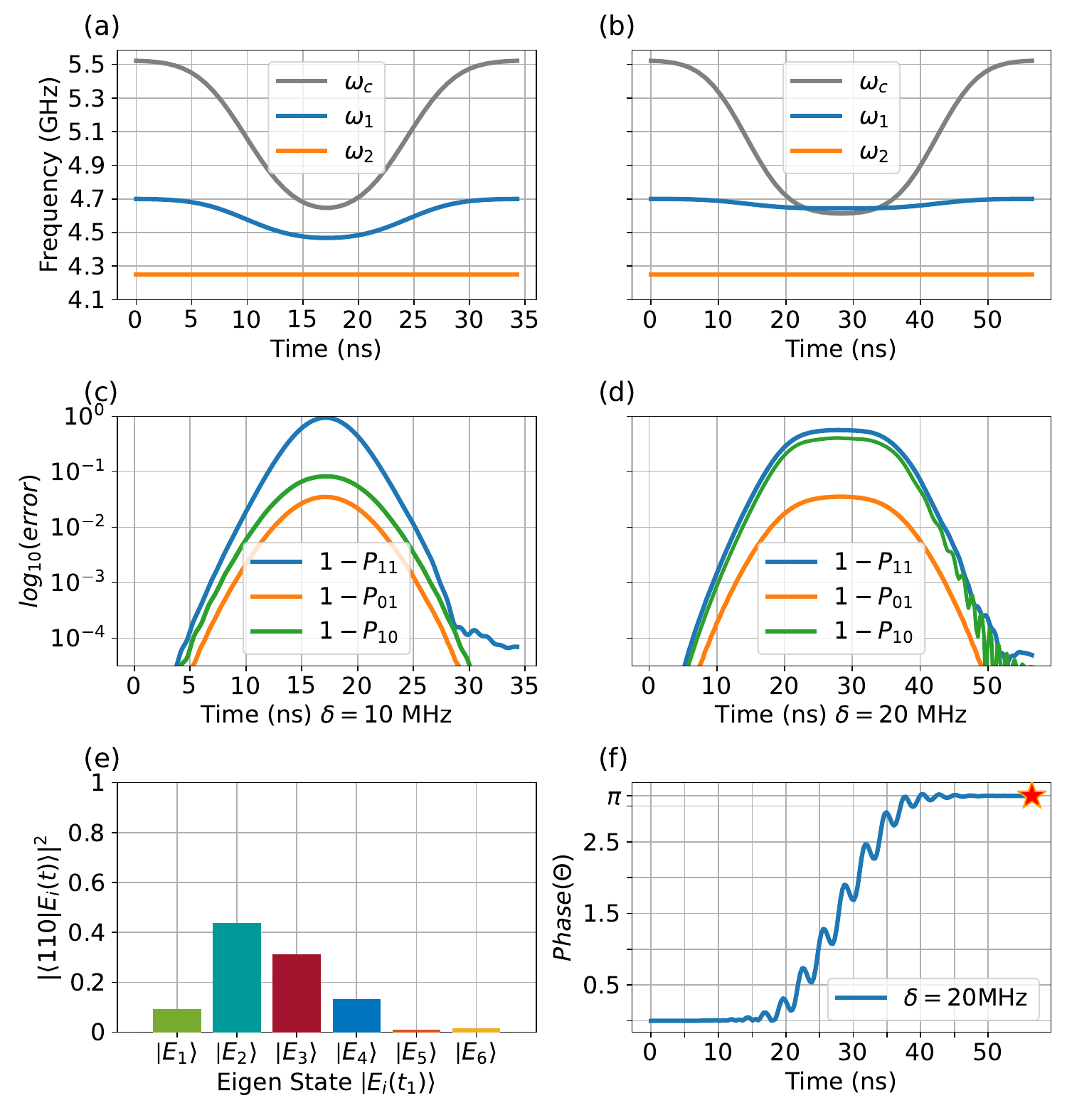}
\caption{\label{wave_pop_delta} (a) Control waveform for the CZ gate at $\delta/2\pi=10~\mathrm{MHz}$. The corresponding leakage and swap errors are shown in (c). (b) Control waveform for the CZ gate at $\delta/2\pi=20~\mathrm{MHz}$. The errors are shown in (d). An error below $10^{-4}$ is still achieved, albeit with an increased gate duration compared to the ideal case ($\delta=0$) in Fig.~\ref{wave_pop_0}. (e), (f) Overlap of the computational basis state $|110\rangle$ with the Hamiltonian eigenstates and the corresponding dynamical phase accumulation during time evolution, plotted as the absolute value, for a qubit anharmonicity mismatch of $\delta = 20~\mathrm{MHz}$.}
\end{figure}

In the architecture shown in Fig.~\ref{wave_pop_delta}(a), we further examined the effect of asymmetric anharmonicity, defined by the offset error $\delta=\mid\alpha_1\mid-\mid\alpha_2\mid$. As shown in Fig.~\ref{wave_pop_delta}, even with substantial asymmetry, such as $\delta/2\pi=10\mathrm{~MHz}$ and $\mathrm{20~MHz,}$ a CZ gate fidelity above 99.99\% remains achievable through optimized pulse shaping, with both leakage and swap errors consistently suppressed below $10^{-4}$, as shown in Fig~\ref{wave_pop_delta}(c),(d). This performance is attained at the cost of a moderate increase in gate duration. At $\delta = 20~\mathrm{MHz}$, the dispersive condition is violated as the coupler frequency traverses the qubit frequencies during the gate operation, as illustrated in Fig.~\ref{wave_pop_delta}(b). By employing the pulse waveform configured for $\delta = 20~\mathrm{MHz}$, we observe a qualitative behavior consistent with that shown in Fig.~\ref{overlap_and_phase_t}: the $|110\rangle$ state is initially driven to non-computational states, expressed as superpositions of Hamiltonian eigenstates, and subsequently refocuses back to $|110\rangle$, thereby successfully completing the intended phase accumulation, as shown in Fig.~\ref{wave_pop_delta}(e),(f).

\begin{figure}[h]
\includegraphics[width=1\linewidth]{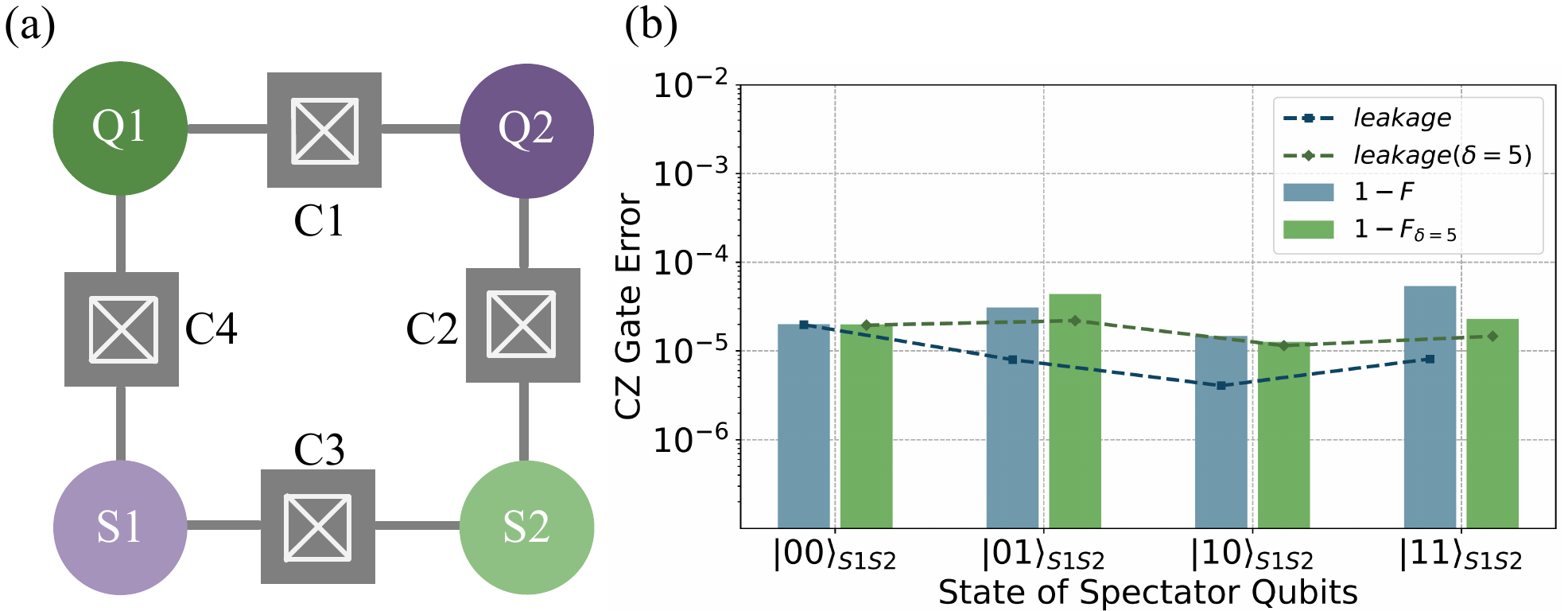}
\caption{\label{spect_error}(a) Four-qubit lattice configuration. Qubits $Q_{1}$ and $Q_{2}$ are used to implement the CZ gate, while $S_{1}$ and $S_{2}$ serve as spectator qubits. The full quantum state of the system is denoted as $|\mathrm{~Q_1,Q_2,S_1,S_2,C_1,C_2,C_3,C_4}\rangle$, where $C_1-C_4$ represent tunable couplers. (b) CZ gate error and leakage error under different spectator qubit states. The blue and green bars represent the gate error for anharmonicity deviations of $\delta = 0$ and $\delta =5$MHz, respectively. The corresponding leakage errors are plotted as lines of the same colors. Notably, for spectator qubits $S_{1}$ and $S_{2}$ initialized in states $|00\rangle,|01\rangle,|10\rangle,\mathrm{~and~}|11\rangle$, the gate error between the active qubits $Q_{1}$ and $Q_{2}$ remains consistently below $10^{-4}$ in all cases.}
\end{figure}

In practice, the presence of spectator qubits is a well-documented source of infidelity in quantum gate operations, primarily mediated by residual interactions. To quantitatively evaluate the effect of spectator qubits, we employ a four-qubit grid architecture in which a CZ gate is performed on two active qubits, while the two spectator qubits are prepared in each of the computational basis states $\left|00\right\rangle$, $\left|01\right\rangle$, $\left|10\right\rangle$, and $\left|11\right\rangle$. As shown in Fig.~\ref{spect_error}(a), the CZ gate is implemented between the gate qubits ($Q_1$, $Q_2$) by frequency tuning of $Q_1$ and $C_1$, while the spectator qubits ($S_1$, $S_2$) and the other couplers are maintained at their idle points. 

The system Hamiltonian can be written as:
\begin{equation}\begin{aligned}\hat{H}&=\sum_{i=1}^{4}\left(\omega_{i}\hat{a}_{i}^{\dagger}\hat{a}_{i}+\frac{\alpha_{i}}{2}\hat{a}_{i}^{\dagger}\hat{a}_{i}^{\dagger}\hat{a}_{i}\hat{a}_{i}+\omega_{ci}\hat{a}_{ci}^{\dagger}\hat{a}_{ci}+\frac{\alpha_{ci}}{2}\hat{a}_{ci}^{\dagger}\hat{a}_{ci}^{\dagger}\hat{a}_{ci}\hat{a}_{ci}\right)\\&+\sum_{i<j}g_{ij}\left(\hat{a}_{i}^{\dagger}\hat{a}_{j}+\hat{a}_{i}\hat{a}_{j}^{\dagger}-\hat{a}_{i}\hat{a}_{j}-\hat{a}_{i}^{\dagger}\hat{a}_{j}^{\dagger}\right),\\&+\sum_{i\leq j}g_{i,cj}\left(\hat{a}_{i}^{\dagger}\hat{a}_{cj}+\hat{a}_{i}\hat{a}_{cj}^{\dagger}-\hat{a}_{i}\hat{a}_{cj}-\hat{a}_{i}^{\dagger}\hat{a}_{cj}^{\dagger}\right),\end{aligned}\end{equation}
where the first line corresponds to the mode Hamiltonian of the qubits and couplers, the second line describes the coupling Hamiltonian between the qubits, and the third line represents the nearest-neighbor interaction Hamiltonian between the qubits and the couplers.

Numerical simulations show that maintaining the coupler between spectator and gate qubits at its idle point allows the CZ gate to consistently achieve an error rate on the order of $10^{-4}$. The gate fidelity for each spectator configuration is detailed in Fig.~\ref{spect_error}(b). Specifically, the blue bars represent the fidelity under ideal anharmonicity conditions. To assess the impact of anharmonicity errors, we also evaluate the CZ gate performance with a qubit anharmonicity deviation of $\delta=5$ MHz, which is depicted by the green bars. Furthermore, the corresponding leakage errors for these two scenarios are superimposed as blue and green dashed lines, respectively. The gate error between the active qubits $Q_1$ and $Q_2$, as well as the leakage error into th remains consistently below $10^{-4}$ in all cases.

\section{Conclusions}
This work presents a fast and high-fidelity implementation of the CZ gate based on a superconducting circuit architecture featuring energy-level-engineered qubits and a tunable coupler. The proposed scheme overcomes the speed limitation of conventional approaches by designing a resonant coupling condition through complementary anharmonicities of a Transmon and an IST qubit. This design enhances the effective qubit-qubit coupling, significantly reducing the duration of gate. Comprehensive numerical simulations demonstrate that the optimized CZ gate achieves $99.99\%$ fidelity within 22 ns, while maintaining error rates below  $10^{-4}$ under practical non-idealities including anharmonicity offsets and spectator qubit interactions. These results show a scalable pathway toward high-performance quantum gates compatible with large-scale quantum processors.

\begin{acknowledgments}
We are grateful to Dr. Peng Zhao from the Quantum Science Center of Guangdong-Hong Kong-Macau Greater Bay Area for his guidance on circuit design, and to Dr. Ji Chu from the Shenzhen Institute for Quantum Science and Engineering for his invaluable advice on waveform optimization. We also extend our sincere thanks to Dr. Mengru Yun from Zhengzhou University for her insightful suggestions on this research scheme. 
\end{acknowledgments}

\appendix
\section{Circuit Hamiltonian }
Fig.~\ref{Circuit}(a) shows the superconducting circuit model of a Transmon qubit and IST qubit coupled to a tunable coupler. The self-capacitance and  Josephson junction energy of qubits/coupler are denoted by $C_{i}$, and $E_{J,i}$ where $i=1,c$.  $E_{J,1}$ and  $E_{J,c}$ are tunable Josephson energy,
\begin{equation}E_{J,i}=E_{J_{i,\Sigma}}\sqrt{\cos^2\left(\frac{\pi\Phi_{e,i}}{\Phi_0}\right)+d_i^2\sin^2\left(\frac{\pi\Phi_{e,i}}{\Phi_0}\right)},\end{equation}
where $\Phi_0=h/2e$ is the flux quantum, $E_{J_{i,\Sigma}}=E_{J_{i,L}}+E_{J_{i,R}}$ is the sum of Josephson energies, $d_{i}=(E_{J_{i,L}}-E_{J_{i,R}})/(E_{J_{i,L}}+E_{J_{i,R}})$ is the junction asymmetry. The Lagrangian of the system can be written as $L=T-U,$ where $T $ and $U$ denote the kinetic energy and the potential energy, respectively. 
\begin{equation}\begin{aligned}{T}&=\frac{1}{2}[C_{1}\dot{\phi}_{1}^{2}+C_{c}\dot{\phi}_{c}^{2}+C_{2}\dot{\phi}_{2}^{2}+C_{1c}(\dot{\phi}_{1}-\dot{\phi}_{c})^{2}\\&+C_{2c}(\dot{\phi}_{2}-\dot{\phi}_{c})^{2}+C_{12}(\dot{\phi}_{1}-\dot{\phi}_{2})^{2}],\end{aligned}\end{equation}
\begin{equation}\begin{aligned}U&=E_{J,1}\cos(\varphi_{1})+E_{J,c}\cos(\varphi_{c})+E_{J,2}\cos(\varphi_{2}+2\pi\phi_{ext})\\&+\frac{1}{2}E_{L}\varphi_{2}^{2},\end{aligned}\end{equation}
The full system Hamiltonian is
\begin{equation}H=\sum_{\varphi_{i}}\dot{\varphi}_{i}\frac{\partial L}{\partial\dot{\varphi}_{i}}-L=\sum_{i=1,2,c}H_{i}+\sum_{i,j=1,2,c}^{i<j}H_{ij},\end{equation}
The Hamiltonians for the Transmon qubit and the coupler are denoted by $H_{1}$  and $H_{c}$, respectively. 
\begin{equation}\begin{gathered}H_{1}=\frac{1}{2}(\frac{1}{2e})^2C_1\dot{\varphi}_1^2-E_{J,1}\cos(\varphi_1)\\=4E_{C,1}n_1^2-E_{J,1}\cos(\varphi_1),\end{gathered}\end{equation}\begin{equation}H_c=4E_{C,c}n_c^2-E_{J,c}\cos(\varphi_c).\end{equation}
$H_{2}$ is the Hamiltonian of IST qubit,
\begin{equation}H_{2}=4E_{C,2}n_{2}^{2}-E_{J2}\cos\left(\varphi_{2}+2\pi\phi_{\mathrm{ext}}\right)+\frac{1}{2}E_{L}\varphi_{L}^{2},\end{equation}
where $E_{J2}$ and $E_{L}$ can be expressed through the equivalent inductances of the junction and the linear inductor as $E_{J2}=\frac{\Phi_0^2}{4\pi^2L_{J2}}$,  $E_{L}=\frac{\Phi_{0}^{2}}{4\pi^{2}L}$. 
The interaction term of Hamiltonian between qubit and coupler is
\begin{equation}H_{ic}=(2e)^{2}\frac{C_{ic}}{C_{i}C_{c}}n_{i}n_{c},\end{equation}
and the direct interaction of qubit is
\begin{equation}H_{12}=\left(2e\right)^{2}\frac{C_{12}+C_{1c}C_{2c}/C_{c}}{C_{1}C_{2}}n_{1}n_{2},\end{equation}

Firstly, we approximate the Hamiltonian of the Transmon qubit and coupler by first expanding the cosine term to leading order and then introducing the standard annihilation and creation operators for the linearized circuit. 
\begin{equation}\begin{aligned}H_{k}&\approx4E_{C,k}n_{k}^{2}-E_{Jk}\left(1-\frac{1}{2}\varphi_{k}^{2}+\frac{1}{24}\varphi_{k}^{4}\right)\\&\approx\omega_{k}a_{k}^{\dagger}a_{k}+\frac{\alpha_{k}}{2}a_{k}^{\dagger}a_{k}(a_{k}^{\dagger}a_{k}-1),\end{aligned}\end{equation}
where \begin{equation}\begin{aligned} & \omega_k=\sqrt{8E_{C,k}E_{J,k}}-E_{C,k},\\  & \alpha_k=-E_{C,k},\\  & \varphi_k=\left(\frac{8E_{C,k}}{E_{J,k}}\right)^{1/4}\frac{a_k+a_k^{\dagger}}{\sqrt{2}},\\  & n_k=\left(\frac{8E_{C,k}}{E_{J,k}}\right)^{-1/4}\frac{a_k-a_k^{\dagger}}{i\sqrt{2}},\end{aligned}\end{equation}
with $k = 1,c$ .

For the IST qubit, we are interested in the regime of a weak inductive shunt with $E_{L}>E_{J,2}$, leading to a strong harmonic confinement and therefore to a single-well phase potential (as opposed to the typical corrugated phase potential of the $E_{L}<<E_{J,2}$ limit). In the half flux quantum $\phi_{\mathrm{ext}}=0.5$, the IST Hamiltonian can be approximately given by
\begin{equation}
\begin{aligned}H_{2}&\approx4E_{C,2}n_{2}^{2}+\frac{1}{2}(E_{L}-E_{J,2})\varphi_{2}^{2}+\frac{1}{24}E_{J,2}\varphi_{2}^{4},
\end{aligned}
\end{equation}
The first transition (qubit) frequency and anharmonicity of IST qubit can be written as:
\begin{equation}\begin{aligned}&\omega_{2}=\sqrt{8E_{C,2}\left(E_{L}-E_{J,2}\right)}+\frac{E_{C,2}E_{J2}}{E_{L}-E_{J2}},\\&\alpha_{2}=\frac{E_{C,2}}{2}\frac{E_J}{E_L-E_{J,2}}.\end{aligned}\end{equation}
At the half-flux-quantum bias point, and taking into account that our numerical simulations retain only the lowest three energy levels, the IST Hamiltonian can be written as:
\begin{equation}H_2\approx\omega_2a_2^\dagger a_2+\frac{\alpha_2}{2}a_2^\dagger a_2(a_2^\dagger a_2-1),\end{equation}
where
\begin{equation}
\begin{aligned}
&\varphi_{2}=\left(\frac{8E_{C,2}}{E_L-E_{J,2}}\right)^{1/4}\frac{a_2+a_2^\dagger}{\sqrt{2}},\\&n_{2}=\left(\frac{8E_{C,2}}{E_L-E_{J,2}}\right)^{-1/4}\frac{a_2-a_2^\dagger}{i\sqrt{2}}
\end{aligned}
\end{equation}
The inter-qubit coupling can be described as
\begin{equation}\begin{aligned}&H_{12}=g_{12}(a_{1}^{\dagger}a_{2}+a_{1}a_{2}^{\dagger}-a_{1}a_{2}-a_{1}^{\dagger}a_{2}^{\dagger}),\\&g_{12}=-(2e)^{2}\frac{C_{12}+C_{1c}C_{2c}/C_{c}}{2C_{1}C_{2}}\left(\frac{8E_{C,2}}{E_{L}-E_{J,2}}\frac{8E_{C,1}}{E_{J,1}}\right)^{-\frac{1}{4}}.\end{aligned}\end{equation}
The coupling of qubit-coupler are
\begin{equation}\begin{aligned}&H_{1c}=g_{1c}(a_{1}^{\dagger}a_{c}+a_{1}a_{c}^{\dagger}-a_{1}a_{c}-a_{1}^{\dagger}a_{c}^{\dagger}),\\&H_{1c}=g_{2c}(a_{2}^{\dagger}a_{c}+a_{2}a_{c}^{\dagger}-a_{2}a_{c}-a_{2}^{\dagger}a_{c}^{\dagger}),\\&g_{1c}=-(2e)^{2}\frac{C_{1c}}{2C_{1}C_{c}}\left(\frac{8E_{C,1}}{E_{J,1}}\frac{8E_{C,c}}{E_{J,c}}\right)^{-\frac{1}{4}},\\g_{2}&_{c}=-(2e)^{2}\frac{C_{2c}}{2C_{2}C_{c}}\left(\frac{8E_{C,2}}{E_{L}-E_{J,2}}\frac{8E_{C,c}}{E_{J,c}}\right)^{-\frac{1}{4}}.\end{aligned}\end{equation}
Therefore the full Hamiltonian of system has following form:
\begin{equation}\begin{aligned}\hat{H}/\hbar & =\sum_{i=1,2,c}\left(\omega_{i}\hat{a}_{i}^{\dagger}\hat{a}_{i}+\frac{\alpha_i}{2}\hat{a}_{i}^{\dagger}\hat{a}_{i}^{\dagger}\hat{a}_{i}\hat{a}_{i}\right)\\  & +\sum_{i<j}g_{ij}\left(\hat{a}_{i}^{\dagger}\hat{a}_{j}+\hat{a}_{i}\hat{a}_{j}^{\dagger}-\hat{a}_{i}\hat{a}_{j}^{}-\hat{a}_{i}^{\dagger}\hat{a}_{j}^{\dagger}\right).\end{aligned}\end{equation}

\section{Cost Function}
This appendix defines the cost function used to optimize the parameters of the CZ gate. The goal is to identify the experimental parameter set $\vec{p}=(t_{\mathrm{hold}},\omega_{c}^{on},\omega_{q}^{on})$ that yields a quantum operation closest to the ideal CZ gate. The proximity to the ideal gate is quantified by a cost function, $C(\vec{p})$, which incorporates two primary error sources: an incorrect conditional phase and leakage out of the computational subspace. The total cost function is defined as the sum of these two error terms:
\begin{equation}C(\vec{p})=C_{\mathrm{phase}}(\vec{p})+C_{\mathrm{leak}}(\vec{p}).\end{equation}
The ideal CZ gate imparts a phase of $\pi$ exclusively to the $|11\rangle$ computational basis state. For a given set of parameters $\vec{p}$, the simulated evolution yields a final state for each basis state. The actual conditional phase, $\theta(\vec{p})$, is extracted from the system's evolution as:
\begin{equation}\theta(\vec{p})=\arg\langle\psi_{_{11}}|\psi_{_f}\rangle-\arg\langle\psi_{_{01}}|\psi_{_f}\rangle-\arg\langle\psi_{_{10}}|\psi_{_f}\rangle+\arg\langle\psi_{_{00}}|\psi_{_f}\rangle,\end{equation}
where $|\psi_f\rangle$ is the final state obtained from evolving the initial state $\sum_{j,k=0,1}\alpha_{jk}\mid jk\rangle$. The phase error is then defined as the squared deviation from the target value
\begin{equation}C_{\mathrm{phase}}(\vec{p})=\left(|\theta(\vec{p})|-\pi\right)^2.\end{equation}

Leakage into non-computational states is a critical source of infidelity. The leakage error is defined as the total population outside the two-qubit computational basis subspace $\{|00\rangle,|01\rangle,|10\rangle,|11\rangle\}$ after the gate operation, averaged over the computational basis states. It is given by
\begin{equation}C_{\mathrm{leak}}(\vec{p})=1-\sum_{j,k=0,1}\alpha_{jk}\left\langle\psi_{f}|ij\right\rangle.\end{equation}
The optimal parameter $\vec{p}_{\mathrm{opt}}$ are  found by numerically minimizing the total cost:
\begin{equation}\vec{p}_{\mathrm{opt}}=\underset{\vec{p}}{\operatorname*{\operatorname*{\operatorname*{\operatorname*{\arg\min}}}}}\{C(\vec{p})\}.\end{equation}
This minimization is performed using a gradient-free algorithm (e.g., Nelder-Mead), which is robust for a low-dimensional parameter space. The numerical simulations of the system's time evolution under the parameterized control pulses were conducted using the QuTiP package~\cite{lambert2024qutip}.
\nocite{*}

{\section{Theoretical Analysis of the CZ Gate for Directly Coupled Qubits}

To analyze the underlying mechanism, we focus on a minimal model of two directly coupled qubits and develop an effective description for the three-level subsystem comprising the states $\{|20\rangle,|11\rangle,|02\rangle\}$ under a finite anharmonicity mismatch condition. We consider a system comprising two capacitively coupled superconducting qubits: a conventional transmon qubit and an inductively shunted transmon (IST) qubit. The system Hamiltonian in the laboratory frame can be expressed as 
\begin{equation}H_{\mathrm{lab}}=\sum_{j=1,2}\omega_ja_j^\dagger a_j+\frac{\alpha_j}{2}a_j^\dagger a_j^\dagger a_ja_j+g_{12}(a_1^\dagger a_2+a_1a_2^\dagger),\end{equation}
where  $\omega_{j}$ denotes the transition frequency of the $j$-th qubit, $\alpha_{j}$ represents the anharmonicity, and $g_{12}$ is the coupling strength between the two qubits. 
In the laboratory frame, the Hamiltonian in the double-excitation subspace $\{|20\rangle,|11\rangle,|02\rangle\}$  is given by
\begin{equation}H_{\mathrm{lab}}=\begin{pmatrix}2\omega_1+\alpha_1&\sqrt{2}g_{12}&0\\\sqrt{2}g_{12}&\omega_1+\omega_2&\sqrt{2}g_{12}\\0&\sqrt{2}g_{12}&2\omega_2+\alpha_2\end{pmatrix},\end{equation}
where $g_{12}$ is the coupling strength between the two qubits in the single-excitation manifold. Moving to a rotating frame and defining the detuning $\Delta=\omega_1-\omega_2$, we obtain 
\begin{equation}H=\begin{pmatrix}\Delta+\alpha_1&\sqrt{2}g_{12}&0\\\sqrt{2}g_{12}&0&\sqrt{2}g_{12}\\0&\sqrt{2}g_{12}&-\Delta+\alpha_2\end{pmatrix}.\end{equation}
In the proposed architecture, the Transmon and IST qubits exhibit opposite-sign anharmonicities, denoted by $\alpha_{1}<0$ and $\alpha_{2}>0$, respectively. To systematically analyze the effect of anharmonicity mismatch, we introduce the sum and difference parameters 
\begin{equation}\delta_+=\frac{\alpha_1+\alpha_2}{2},\quad\delta_-=\frac{\alpha_1-\alpha_2}{2}.\end{equation}
Defining a shifted detuning parameter $\Delta^{\prime}=\Delta+\delta_{-}$, the Hamiltonian can be simplified to 
\begin{equation}H^{\prime}=\begin{pmatrix}\Delta^{\prime}&\sqrt2g_{12}&0\\\sqrt2g_{12}&-\delta_+&\sqrt2g_{12}\\0&\sqrt2g_{12}&-\Delta^{\prime}\end{pmatrix}.\end{equation}
Setting $\Delta^{\prime}=0$ yields the resonance condition $\Delta+\delta_-=0$. Under this condition, the Hamiltonian reduces to 
\begin{equation}H^{\prime}|_{\Delta^{\prime}=0}=\begin{pmatrix}0&\sqrt{2}g_{12}&0\\\sqrt{2}g_{12}&-\delta_{+}&\sqrt{2}g_{12}\\0&\sqrt{2}g_{12}&0\end{pmatrix}.\end{equation}
The symmetric structure of this Hamiltonian admits a dark state of the form $|D\rangle=\frac{1}{\sqrt{2}}\left(|20\rangle-|02\rangle\right),$ which is decoupled from the $\left|11\right\rangle$ state. This can be verified by direct calculation $H^{\prime}|D\rangle=0.$ Notably, even in the presence of anharmonicity mismatch, one can tune the qubit frequencies to enforce 
$\Delta^{\prime}=0$, in which case the dark state remains decoupled from the system dynamics. Correspondingly, we define the bright state as $|B\rangle=\frac{1}{\sqrt{2}}\left(|20\rangle+|02\rangle\right).$
In the basis $\{|11\rangle,|B\rangle,|D\rangle\}$, the Hamiltonian $H^{\prime}$ takes the form 
\begin{equation}H^{\prime}=\begin{pmatrix}-\delta_+&{2}g_{12}&0\\{2}g_{12}&0&0\\0&0&0\end{pmatrix}\quad\mathrm{for}~ \Delta^{\prime}=0.\end{equation}

\begin{figure}[]
\includegraphics[width=0.9\linewidth]{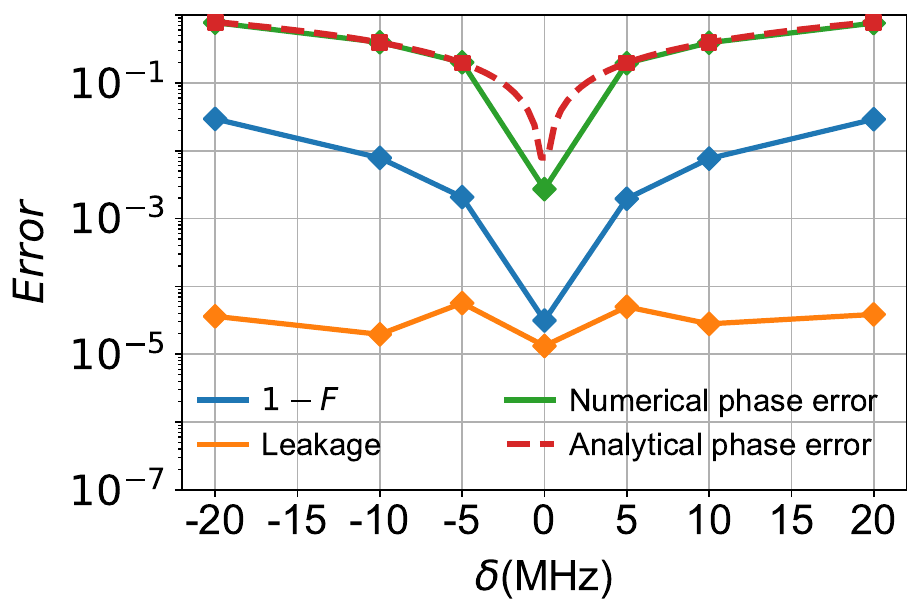}
\caption{\label{Delta_error_no_cpl} 
Error contributions to the CZ gate as functions of the anharmonicity offset in a directly capacitively coupled Transmon–IST system. The plotted metrics include the total gate error (infidelity $1-F$), leakage error, and phase shift error $\Delta\phi$ (in radians), defined as $\Delta \phi  = \left| {{\phi _{ZZ}} - \pi } \right|$.  Both the analytical expression Eq.~(C11) and numerical simulations confirm that phase shift error constitutes the dominant error source under anharmonicity mismatch. 
}
\end{figure}

Since the dark state  $|D\rangle$ is completely decoupled, the dynamics within the computational subspace is governed by the effective two-level Hamiltonian 
\begin{equation}H_{\mathrm{eff}}=\begin{pmatrix}-\delta_+&2g_{12}\\2g_{12}&0\end{pmatrix}.\end{equation}
The time evolution operator in the $\{|11\rangle,|B\rangle\}$ basis can be expressed as 
\begin{equation}\begin{aligned}U(t)&=e^{-iH_{\mathrm{eff}}t}\\&=e^{\frac{i\delta_{+}t}{2}}\left(\begin{array}{cc}\cos\frac{\Omega t}{2}+i\cos\theta\sin\frac{\Omega t}{2}&-i\sin\theta\sin\frac{\Omega t}{2}\\-i\sin\theta\sin\frac{\Omega t}{2}&\cos\frac{\Omega t}{2}-i\cos\theta\sin\frac{\Omega t}{2}\end{array}\right),\end{aligned}\end{equation}
where $\Omega=\sqrt{\delta_+^2+16g^2}$ is the generalized Rabi frequency, and the mixing angle $\theta$ is defined by 
$\sin\theta=\frac{4g_{12}}{\Omega},\quad\cos\theta=\frac{\delta_+}{\Omega}.$ It can be seen that in the presence of a $\delta_{+}$ error, the system exhibits an inherent trade‑off: enforcing a conditional phase of $\pi$ on the $|11\rangle$ state amplifies leakage because the population does not fully return to $|11\rangle$. Conversely, suppressing leakage so that the state returns cleanly to $|11\rangle$ results in a significant residual phase error. 
For an initial state $|\psi(0)\rangle=|11\rangle$, the population in the $\left|11\right\rangle$ state evolves according to the standard two-level formula \begin{equation}P_{11}(t)=|\langle11|e^{-iH_\mathrm{eff}t}|11\rangle|^2=1-\frac{16g_{12}^2}{\Omega^2}\mathrm{sin}^2\left(\frac{\Omega t}{2}\right).\end{equation}
In the limit $\delta_+\ll g_{12}$, we have $\Omega\approx4g_{12}$, and the population exhibits complete Rabi oscillations between $\left|11\right\rangle$ and $\left|B\right\rangle$ with period  $t=\pi/2g_{12}$. 
We introduce the small parameter $\epsilon=\delta_+/(4g)$ to perform a perturbative expansion. To first order in $\epsilon$, the accumulated phase error on $|11\rangle$ is 
\begin{equation}\Delta\phi\approx\frac{\pi\delta_+}{4g_{12}}.\end{equation}

Building on the analytical results from the preceding analysis, which identify phase error as the dominant source of infidelity for the CZ gate under anharmonicity offsets, we proceed to numerically quantify the key error metrics—specifically, the total gate infidelity, leakage error, and conditional phase error—across a range of offset values.
The frequency pulse profile that we use throughout this work is based on Eq.~(12).
\begin{figure}[]
\includegraphics[width=1\linewidth]{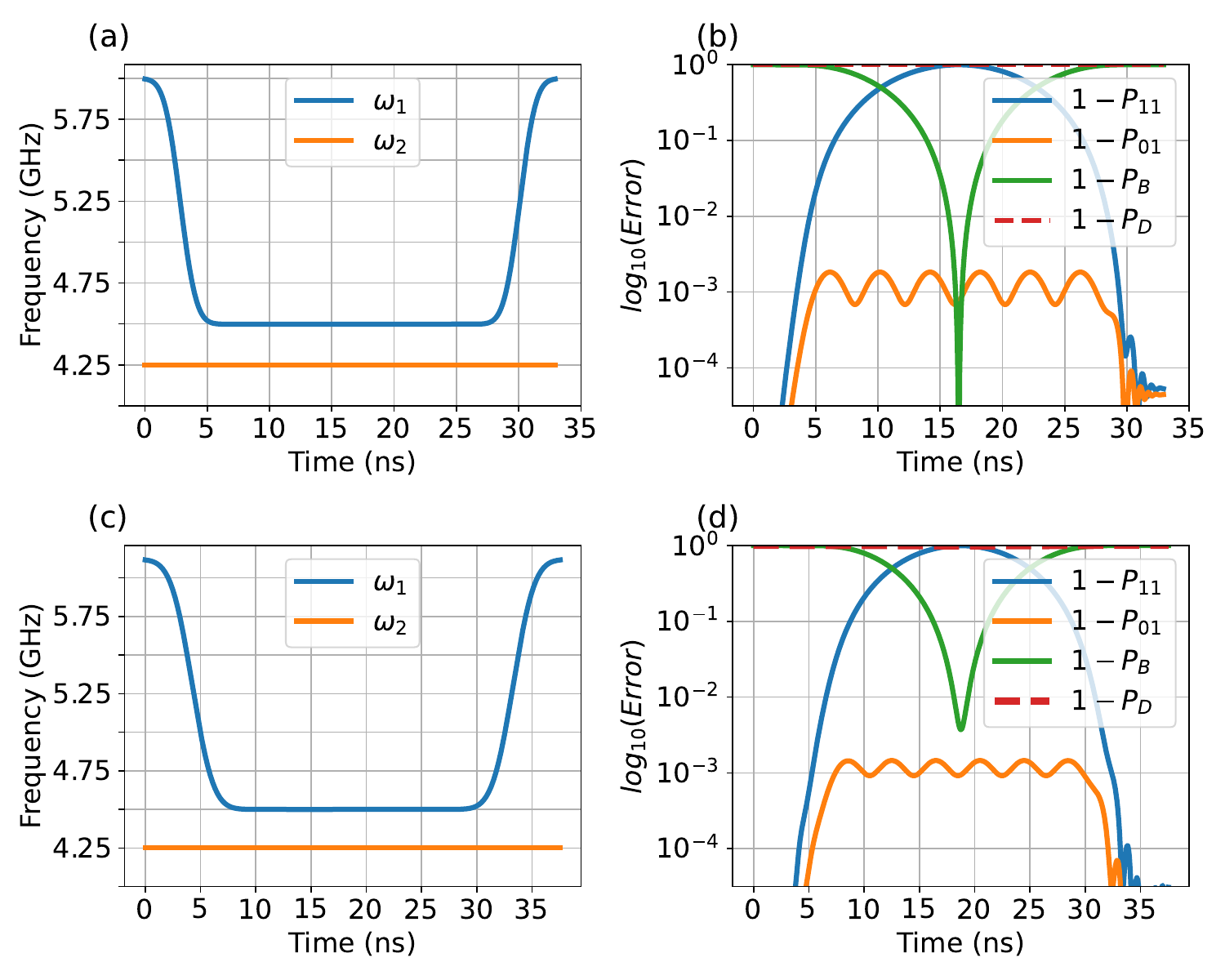}
\caption{\label{no_cpl_population} Control waveforms and population dynamics for a CZ gate in a directly coupled two-qubit system. (a), (b) The control pulse and the corresponding time evolution of the state populations for the case of zero anharmonicity error ($\delta=0$). (c), (d) Analogous results for the case with an anharmonicity error of $\delta=5$ MHz.}
\end{figure}

Consistent with the preceding analytical findings, our numerical results confirm that conditional phase accumulation constitutes the dominant error source under anharmonicity offsets, as illustrated in Fig.~\ref{Delta_error_no_cpl}, while leakage remains consistently suppressed below $10^{-4}$. Furthermore, the numerically obtained phase error agrees well with the analytical prediction of Eq.~(C11).

To concretely illustrate these dynamics, we initialize the system in the $|11\rangle$ and $|01\rangle$ states and compute the leakage error ($\varepsilon_{\mathrm{leak}} = 1 - P_{11}$), the SWAP error ($\varepsilon_{\mathrm{swap}} = 1 - P_{01}$), and the population transfers to the superposition states $|B\rangle$ and $|D\rangle$. We compare the evolution under the ideal condition ($\delta=0$) and a representative anharmonicity offset of $\delta=5~\mathrm{MHz}$. In the ideal case, a complete coherent exchange is observed between the $|11\rangle$ and $|B\rangle$ states, as depicted in Fig.~\ref{no_cpl_population}(b). In the presence of an anharmonicity mismatch, however, we find that while leakage from the $|11\rangle$ state can be mitigated by fine-tuning the qubit frequencies (Fig.~\ref{no_cpl_population}(d)), this suppression comes at the cost of a significant residual phase error. Consequently, the gate fidelity degrades by approximately an order of magnitude, as shown in Fig.~\ref{Delta_error_no_cpl}. This trade-off highlights a fundamental constraint: the finite anharmonicity primarily induces a phase shift that scales linearly with $\delta_{+}$, while leakage remains negligible. This limitation motivates the introduction of a tunable coupler, which provides an additional degree of freedom to effectively decouple these competing error sources.

\bibliography{Appendix}

@article{acharya2024quantum,
  title={Quantum error correction below the surface code threshold},
  author={Acharya, Rajeev and Abanin, Dmitry A and Aghababaie-Beni, Laleh and Aleiner, Igor and Andersen, Trond I and Ansmann, Markus and Arute, Frank and Arya, Kunal and Asfaw, Abraham and Astrakhantsev, Nikita and others},
  journal={Nature},
  year={2024}
}

@article{klimov2024optimizing,
  title={Optimizing quantum gates towards the scale of logical qubits},
  author={Klimov, Paul V and Bengtsson, Andreas and Quintana, Chris and Bourassa, Alexandre and Hong, Sabrina and Dunsworth, Andrew and Satzinger, Kevin J and Livingston, William P and Sivak, Volodymyr and Niu, Murphy Yuezhen and others},
  journal={Nature Communications},
  volume={15},
  number={1},
  pages={2442},
  year={2024},
  publisher={Nature Publishing Group UK London}
}

@article{gyenis2021experimental,
  title={Experimental realization of a protected superconducting circuit derived from the 0--$\pi$ qubit},
  author={Gyenis, Andr{\'a}s and Mundada, Pranav S and Di Paolo, Agustin and Hazard, Thomas M and You, Xinyuan and Schuster, David I and Koch, Jens and Blais, Alexandre and Houck, Andrew A},
  journal={PRX Quantum},
  volume={2},
  number={1},
  pages={010339},
  year={2021},
  publisher={APS}
}

@article{siddiqi2021engineering,
  title={Engineering high-coherence superconducting qubits},
  author={Siddiqi, Irfan},
  journal={Nature Reviews Materials},
  volume={6},
  number={10},
  pages={875--891},
  year={2021},
  publisher={Nature Publishing Group UK London}
}

@article{gyenis2021moving,
  title={Moving beyond the transmon: Noise-protected superconducting quantum circuits},
  author={Gyenis, Andr{\'a}s and Di Paolo, Agustin and Koch, Jens and Blais, Alexandre and Houck, Andrew A and Schuster, David I},
  journal={PRX Quantum},
  volume={2},
  number={3},
  pages={030101},
  year={2021},
  publisher={APS}
}

@article{hassani2023inductively,
  title={Inductively shunted transmons exhibit noise insensitive plasmon states and a fluxon decay exceeding 3 hours},
  author={Hassani, Farid and Peruzzo, Matilda and Kapoor, LN and Trioni, Andrea and Zemlicka, Martin and Fink, Johannes M},
  journal={Nature Communications},
  volume={14},
  number={1},
  pages={3968},
  year={2023},
  publisher={Nature Publishing Group UK London}
}

@article{somoroff2023millisecond,
  title={Millisecond coherence in a superconducting qubit},
  author={Somoroff, Aaron and Ficheux, Quentin and Mencia, Raymond A and Xiong, Haonan and Kuzmin, Roman and Manucharyan, Vladimir E},
  journal={Physical Review Letters},
  volume={130},
  number={26},
  pages={267001},
  year={2023},
  publisher={APS}
}

@article{place2021new,
  title={New material platform for superconducting transmon qubits with coherence times exceeding 0.3 milliseconds},
  author={Place, Alexander PM and Rodgers, Lila VH and Mundada, Pranav and Smitham, Basil M and Fitzpatrick, Mattias and Leng, Zhaoqi and Premkumar, Anjali and Bryon, Jacob and Vrajitoarea, Andrei and Sussman, Sara and others},
  journal={Nature communications},
  volume={12},
  number={1},
  pages={1779},
  year={2021},
  publisher={Nature Publishing Group UK London}
}

@article{wang2022towards,
  title={Towards practical quantum computers: Transmon qubit with a lifetime approaching 0.5 milliseconds},
  author={Wang, Chenlu and Li, Xuegang and Xu, Huikai and Li, Zhiyuan and Wang, Junhua and Yang, Zhen and Mi, Zhenyu and Liang, Xuehui and Su, Tang and Yang, Chuhong and others},
  journal={npj Quantum Information},
  volume={8},
  number={1},
  pages={3},
  year={2022},
  publisher={Nature Publishing Group UK London}
}

@article{ganjam2024surpassing,
  title={Surpassing millisecond coherence in on chip superconducting quantum memories by optimizing materials and circuit design},
  author={Ganjam, Suhas and Wang, Yanhao and Lu, Yao and Banerjee, Archan and Lei, Chan U and Krayzman, Lev and Kisslinger, Kim and Zhou, Chenyu and Li, Ruoshui and Jia, Yichen and others},
  journal={Nature Communications},
  volume={15},
  number={1},
  pages={3687},
  year={2024},
  publisher={Nature Publishing Group UK London}
}

@article{verjauw2022path,
  title={Path toward manufacturable superconducting qubits with relaxation times exceeding 0.1 ms},
  author={Verjauw, J and Acharya, R and Van Damme, J and Ivanov, Ts and Lozano, D Perez and Mohiyaddin, FA and Wan, D and Jussot, J and Vadiraj, AM and Mongillo, M and others},
  journal={npj Quantum Information},
  volume={8},
  number={1},
  pages={93},
  year={2022},
  publisher={Nature Publishing Group UK London}
}

@article{kim2021enhanced,
  title={Enhanced coherence of all-nitride superconducting qubits epitaxially grown on silicon substrate},
  author={Kim, Sunmi and Terai, Hirotaka and Yamashita, Taro and Qiu, Wei and Fuse, Tomoko and Yoshihara, Fumiki and Ashhab, Sahel and Inomata, Kunihiro and Semba, Kouichi},
  journal={Communications Materials},
  volume={2},
  number={1},
  pages={98},
  year={2021},
  publisher={Nature Publishing Group UK London}
}

@article{bland20252d,
  title={2D transmons with lifetimes and coherence times exceeding 1 millisecond},
  author={Bland, Matthew P and Bahrami, Faranak and Martinez, Jeronimo GC and Prestegaard, Paal H and Smitham, Basil M and Joshi, Atharv and Hedrick, Elizabeth and Pakpour-Tabrizi, Alex and Kumar, Shashwat and Jindal, Apoorv and others},
  journal={arXiv preprint arXiv:2503.14798},
  year={2025}
}

@article{sung2021realization,
  title={Realization of high-fidelity CZ and ZZ-free iSWAP gates with a tunable coupler},
  author={Sung, Youngkyu and Ding, Leon and Braum{\"u}ller, Jochen and Veps{\"a}l{\"a}inen, Antti and Kannan, Bharath and Kjaergaard, Morten and Greene, Ami and Samach, Gabriel O and McNally, Chris and Kim, David and others},
  journal={Physical Review X},
  volume={11},
  number={2},
  pages={021058},
  year={2021},
  publisher={APS}
}

@article{martinis2014fast,
  title={Fast adiabatic qubit gates using only $\sigma$ z control},
  author={Martinis, John M and Geller, Michael R},
  journal={Physical Review A},
  volume={90},
  number={2},
  pages={022307},
  year={2014},
  publisher={APS}
}

@article{barends2019diabatic,
  title={Diabatic gates for frequency-tunable superconducting qubits},
  author={Barends, Rami and Quintana, CM and Petukhov, AG and Chen, Yu and Kafri, Dvir and Kechedzhi, Kostyantyn and Collins, Roberto and Naaman, Ofer and Boixo, Sergio and Arute, F and others},
  journal={Physical review letters},
  volume={123},
  number={21},
  pages={210501},
  year={2019},
  publisher={APS}
}

@article{xu2020high,
  title={High-fidelity, high-scalability two-qubit gate scheme for superconducting qubits},
  author={Xu, Yuan and Chu, Ji and Yuan, Jiahao and Qiu, Jiawei and Zhou, Yuxuan and Zhang, Libo and Tan, Xinsheng and Yu, Yang and Liu, Song and Li, Jian and others},
  journal={Physical review letters},
  volume={125},
  number={24},
  pages={240503},
  year={2020},
  publisher={APS}
}

@article{li2019realisation,
  title={Realisation of high-fidelity nonadiabatic CZ gates with superconducting qubits},
  author={Li, Shaowei and Clark, Juno and Wang, Shiyu and Wu, Yulin and Gong, Ming and Yan, Zhiguang and Rong, Hao and Deng, Hui and Zha, Chen and Guo, Cheng and others},
  journal={npj Quantum Information},
  volume={5},
  number={1},
  pages={84},
  year={2019},
  publisher={Nature Publishing Group UK London}
}

@article{chu2021coupler,
  title={Coupler-assisted controlled-phase gate with enhanced adiabaticity},
  author={Chu, Ji and Yan, Fei},
  journal={Physical Review Applied},
  volume={16},
  number={5},
  pages={054020},
  year={2021},
  publisher={APS}
}

@article{goto2022double,
  title={Double-transmon coupler: Fast two-qubit gate with no residual coupling for highly detuned superconducting qubits},
  author={Goto, Hayato},
  journal={Physical review applied},
  volume={18},
  number={3},
  pages={034038},
  year={2022},
  publisher={APS}
}

@article{huang2024fast,
  title={Fast ZZ-free entangling gates for superconducting qubits assisted by a driven resonator},
  author={Huang, Ziwen and Kim, Taeyoon and Roy, Tanay and Lu, Yao and Romanenko, Alexander and Zhu, Shaojiang and Grassellino, Anna},
  journal={Physical Review Applied},
  volume={22},
  number={3},
  pages={034007},
  year={2024},
  publisher={APS}
}

@article{jiang2025microwave,
  title={Microwave-activated two-qubit gates for fixed-coupling and fixed-frequency transmon qubits},
  author={Jiang, Ling and Xu, Peng and Wu, Shengjun and Sun, Jian-An and Dou, Fu-Quan},
  journal={Physical Review A},
  volume={111},
  number={3},
  pages={032609},
  year={2025},
  publisher={APS}
}

@article{zhao2020high,
  title={High-contrast zz interaction using superconducting qubits with opposite-sign anharmonicity},
  author={Zhao, Peng and Xu, Peng and Lan, Dong and Chu, Ji and Tan, Xinsheng and Yu, Haifeng and Yu, Yang},
  journal={Physical Review Letters},
  volume={125},
  number={20},
  pages={200503},
  year={2020},
  publisher={APS}
}

@article{ku2020suppression,
  title={Suppression of unwanted ZZ interactions in a hybrid two-qubit system},
  author={Ku, Jaseung and Xu, Xuexin and Brink, Markus and McKay, David C and Hertzberg, Jared B and Ansari, Mohammad H and Plourde, BLT},
  journal={Physical review letters},
  volume={125},
  number={20},
  pages={200504},
  year={2020},
  publisher={APS}
}

@article{yan2018tunable,
  title={Tunable coupling scheme for implementing high-fidelity two-qubit gates},
  author={Yan, Fei and Krantz, Philip and Sung, Youngkyu and Kjaergaard, Morten and Campbell, Daniel L and Orlando, Terry P and Gustavsson, Simon and Oliver, William D},
  journal={Physical Review Applied},
  volume={10},
  number={5},
  pages={054062},
  year={2018},
  publisher={APS}
}

@article{zhao2022quantum,
  title={Quantum crosstalk analysis for simultaneous gate operations on superconducting qubits},
  author={Zhao, Peng and Linghu, Kehuan and Li, Zhiyuan and Xu, Peng and Wang, Ruixia and Xue, Guangming and Jin, Yirong and Yu, Haifeng},
  journal={PRX quantum},
  volume={3},
  number={2},
  pages={020301},
  year={2022},
  publisher={APS}
}

@article{li2024realization,
  title={Realization of high-fidelity CZ gate based on a double-transmon coupler},
  author={Li, Rui and Kubo, Kentaro and Ho, Yinghao and Yan, Zhiguang and Nakamura, Yasunobu and Goto, Hayato},
  journal={Physical Review X},
  volume={14},
  number={4},
  pages={041050},
  year={2024},
  publisher={APS}
}

@article{xu2021zz,
  title={ZZ freedom in two-qubit gates},
  author={Xu, Xuexin and Ansari, MH},
  journal={Physical review applied},
  volume={15},
  number={6},
  pages={064074},
  year={2021},
  publisher={APS}
}

@article{yan2016flux,
  title={The flux qubit revisited to enhance coherence and reproducibility},
  author={Yan, Fei and Gustavsson, Simon and Kamal, Archana and Birenbaum, Jeffrey and Sears, Adam P and Hover, David and Gudmundsen, Ted J and Rosenberg, Danna and Samach, Gabriel and Weber, Steven and others},
  journal={Nature communications},
  volume={7},
  number={1},
  pages={12964},
  year={2016},
  publisher={Nature Publishing Group UK London}
}

@article{fasciati2024complementing,
  title={Complementing the transmon by integrating a geometric shunt inductor},
  author={Fasciati, Simone D and Shteynas, Boris and Campanaro, Giulio and Bakr, Mustafa and Cao, Shuxiang and Chidambaram, Vivek and Wills, James and Leek, Peter J},
  journal={arXiv preprint arXiv:2410.10416},
  year={2024}
}

@article{krinner2020benchmarking,
  title={Benchmarking coherent errors in controlled-phase gates due to spectator qubits},
  author={Krinner, Sebastian and Lazar, Stefania and Remm, Ants and Andersen, Christian K and Lacroix, Nathan and Norris, Graham J and Hellings, Christoph and Gabureac, Mihai and Eichler, Christopher and Wallraff, Andreas},
  journal={Physical Review Applied},
  volume={14},
  number={2},
  pages={024042},
  year={2020},
  publisher={APS}
}

@article{pedersen2007fidelity,
  title={Fidelity of quantum operations},
  author={Pedersen, Line Hjortsh{\o}j and M{\o}ller, Niels Martin and M{\o}lmer, Klaus},
  journal={Physics Letters A},
  volume={367},
  number={1-2},
  pages={47--51},
  year={2007},
  publisher={Elsevier}
}

@article{lambert2024qutip,
  title={Qutip 5: The quantum toolbox in python},
  author={Lambert, Neill and Gigu{\`e}re, Eric and Menczel, Paul and Li, Boxi and Hopf, Patrick and Su{\'a}rez, Gerardo and Gali, Marc and Lishman, Jake and Gadhvi, Rushiraj and Agarwal, Rochisha and others},
  journal={arXiv preprint arXiv:2412.04705},
  year={2024}
}

\end{document}